\documentclass[12pt]{article}
\usepackage{latexsym,graphicx,multirow}
\usepackage{float}
\usepackage{amssymb}
\usepackage{amsmath}
\usepackage{amscd}
\usepackage{amsthm}
\usepackage[left=2cm,top=2.5cm,right=2.5cm,bottom=1.5cm]{geometry}
\usepackage{hyperref}
\usepackage{epstopdf}

\begin{document}
	
	\begin{center}
	\large{\bf{An Axially Symmetric Transitioning models with Observational Constraints}} \\
	\vspace{10mm}
	\normalsize{ Vinod Kumar Bhardwaj$^1$,  Archana Dixit$^2$, Rita Rani$^3$, G. K. Goswami$^4$, Anirudh Pradhan$^5$ }\\
	\vspace{5mm}
	\normalsize{$^{1, 2}$Department of Mathematics, Institute of Applied Sciences and Humanities, GLA University, Mathura-281 406, Uttar Pradesh, India}\\
	\vspace{2mm}
	\normalsize{$^{3,4}$Department of Mathematics, Netaji Subhas University of Technology, Delhi, India}\\
	\vspace{5mm}
	\normalsize{$^{5}$Centre for Cosmology, Astrophysics and Space Science (CCASS), GLA University, Mathura-281 406, Uttar Pradesh, India}\\
	\vspace{2mm}
	$^1$E-mail: dr.vinodbhardwaj@gmail.com \\
	\vspace{2mm}
	$^2$E-mail:archana.dixit@gla.ac.in\\
	\vspace{2mm}
	$^3$E-mail: rita.ma19@nsut.ac.in \\
	\vspace{2mm}
	$^4$E-mail: gk.goswami9@gmail.com \\
	\vspace{2mm}
	$^5$E-mail:pradhan.anirudh@gmail.com\\
	
	\vspace{10mm}
	
	%\date{}
	%\maketitle
\end{center}

\begin{abstract}
 In this study, we have demonstrated the expansion history of an axially symmetric Bianchi type-I model of the universe. Our model as of now presents an accelerating universe, which had been in the decelerating phase in the past. Roles of the two crucial Hubble~$H(z)$ and deceleration~$q(z)$ parameters are examined. The energy parameters of the universe are estimated with the help of the latest observational Hubble data (46-data points) and Pantheon data (the latest compilation of SNIa with 40 binned in the redshift range $0.014 \leq z \leq 1.62)$. We also discuss the stability analysis of the model by state finder diagnosis. The analysis reveals that in late time, the model is a quintessence type and points towards the $\Lambda$CDM model. Our developed model agrees with observational findings in a proper way. We have discussed some of the physical aspects of the model.
		
\end{abstract}

\smallskip 
{\bf Keywords} : LRS Bianchi-I, $ \Lambda$CDM Model, Statefinder  diagnosis.

	PACS: 98.80.-k \\
	
	%%%%%%%%%%%%%%%%%%%%%%%%%%%%%%%%%%%%%%%%%%%%%%%%%% Section 1 %%%%%%%%%%%%%%%%%%%%%%%%%%
\section{Introduction}

Type Ia Supernova observations, announced in 1998, indicate that the physical universe is accelerating which had been in the decelerating phase in the past\cite{ref1,ref2}. The background behind these observations are the studies of supernova exploration and explosions. SN Ia supernovae are standard candles (populations II and III stars). At shorter redshift, nearby SN Ia help in the determination of Hubble constant H whereas large red-shifted long distant objects provide an estimation of deceleration constant q \cite{ref3, ref4,ref5}.  The Supernova Cosmology Project was launched in 1988, in which magnitude-redshift relations were used to estimate the cosmological parameters. In this project, cosmological parameters were estimated with the help of more than 75 SN Ia ($0.18\leq z \leq 0.86$) magnitude-redshift relation.  Authors\cite{ref6} discussed that the cosmological constant-based dark energy is dominant over baryon matter-energy. Authors \cite{ref7} worked with the 33 additional high-red shifted supernovae and developed a confidence region of an accelerating universe. They have also searched for a possibility of a  low-mass $\Lambda =0$ cosmology.  In this context, Riess et al. \cite{ref8} had also done an independent study of 10 high-redshift supernovae and found that the universe is accelerating. In this direction, CMB anisotropy \cite{ref9}, SN Ia  magnitude-redshift relation \cite{ref10,ref11}, baryonic acoustic oscillation (BAO), peak length scale  \cite{ref12,ref13,ref14,ref15}, and Hubble parameter versus redshift measurements \cite{ref16,ref17,ref18,ref19} point to the accelerating universe.\\

Most recent, Hubble parameters for various red-shifts have been determined with the help of cosmic chronometric (CC) and  BAO techniques \cite{ref16,ref19,ref20}. They provided cosmic ``deceleration-acceleration transition" when the universe is thought to have the presence of dark energy in abundance along with non-relativistic baryon matter. This redshift  transition were discussed in \cite{ref19,ref21,ref22}. Hubble H(z) data sets are also used to derive energy parameters $\Omega_m$ and $\Omega_{de}$ \cite{ref23,ref24,ref25,ref26,ref27,ref28,ref29,ref30} and are comparable with the results with SNIa data, BAO and CMB observations. They estimated the present value of Hubble constant $H_{0}$ \cite{ref31,ref32}. The derived values of $H_{0}$ are compatible with Huchra's $H_{0}$ compilation statistical analysis \cite{ref17}. All of the above findings were discovered by a simple analysis of ``55 supernova data points". \\

46 Hubble parameters H(z) data set in the  redshifts range $0\leq z\leq 2.36$ were used to estimate present value of Hubble constant along with baryon and dark  energy parameters \cite{ref33,ref34,ref35,ref36}. The authors of these references have also analyzed the cumulative effect of  H(z) data, SN Ia pantheon compilation data, and BAO data in the estimation of model parameters. They have used the $\chi^{2}$ minimization technique. SN Ia pantheon compilation data set is given in Scolnic et al. \cite{ref37}.\\

 The late-time acceleration of the universe is described in General relativity (GR) by dark energy density along with matter density in Einstein's field equation \cite{ref38,ref39,ref40,ref41}. It is suggested that dark energy of repulsive anti gravitating origin present in abundance in our universe, which is responsible for the change in the mode of the universe.  In some modified theories of gravity accelerated expansion of the universe has been explained without applying the DE component in a different way of thinking. A variety of gravity theories such as f(R) \cite{ref42,ref43,ref44}, f(T) \cite{ref45}, and f(G) \cite {ref46}, f(Q,T) \cite {ref47,ref48}, f(Q) \cite {ref49,ref50,ref51}, f(R,G) \cite {ref52}, f(T,B) gravity \cite {ref53}, and Einstein-Gauss-Bonnet theory \cite {ref54,ref55,ref56,ref57,ref58,ref59} were proposed. In the same direction, Harko et al. \cite{ref60} developed f(R,T)gravity theory. In these modified theories, non-linear curvature scalar and energy traces describe acceleration in the universe. Their origin lies in the corrections towards curvature and energy trace. \\
 
Off late considerable number of Bianchi type-I cosmological models that fit well with the latest observations and exhibit acceleration in the universe were surfaced
in the literature \cite{ref61,ref62,ref63,ref64,ref65,ref66}. It is a well-known fact that neutrino viscosity develops anisotropy in the universe and was in abundance during the primordial fireball(P.F.) \cite{ref67,ref68}. As of now the contribution of neutrinos and CMBR(Cosmic Microwave Background Radiations) in the total content of the universe are ~0.1\% $-$ 0.5\% \cite{ref69}. They are leftover remains of the P.F. . So just after the investigation of the CMBR phenomenon, Spatially homogeneous and anisotropic cosmological models attracted the scientific community \cite{ref70,ref71}. It is 1962, when Bianchi type-I cosmological models were introduced in the book ``An Introduction to Current Research" \cite{ref70}. Sometimes they were called Heckmann and Schucking models in the author's name. The Bianchi-type models are the best and simplest anisotropic models, which completely describe the anisotropic effects. The advantages of these anisotropic models are that they play an important role in describing the history of the early Universe and that they contribute to the formation of more generalized cosmological models than isotropic FRW models \cite{ref72,ref73,ref74}.\\
 
Locally rotational symmetric(LRS) spacetimes are a sub-class of Bianchi type I models in the sense that they are spatially homogeneous and describe symmetry along a particular direction(axis). They may also be called axially symmetric spacetimes. A considerable number of present days accelerating universe models on LRS spacetimes have appeared in the literatures  \cite{ref75,ref76,ref77,ref78,ref79,ref80,ref81}. \\
 
  In this study, we have demonstrated the expansion history of an axially symmetric Bianchi type-I model of the universe. Our model as of now presents an accelerating universe, which had been in the decelerating phase in the past. Roles of the two crucial Hubble~$H(z)$ and deceleration~$q(z)$ parameters are examined. The energy parameters of the universe are estimated with the help of the latest observational Hubble data (46-data points) and Pantheon data (the latest compilation of SNIa with 40 binned in the redshift range $0.014 \leq z \leq 1.62)$. We also discuss the stability analysis of the model by state finder diagnosis. The analysis reveals that in late time, the model is a quintessence type and points towards the $\Lambda$ CDM model. Our developed model agrees with observational findings in a proper way. We have discussed some of the physical aspects of the model.\\
 	
 The paper is structured as follows: The present cosmological scenario is discussed in section I. In section 2 we present a model and field equation. In section 3, we constrain the model parameters using Hubble data set H(z) and Pantheon data set. We have also discussed the Luminosity Distance,  Apparent Magnitude, and Distance Modulus of the model.  Section 4 contains the age of the universe. Section 5 shows the behavior of the deceleration parameter. Evolutionary trajectories are discussed in section 6. Conclusions are mentioned in section 7.
 	
 	%%%%%%%%%%%%%%%%%%%%%%%%%%%%%%%%%%%%%%%%%% Section 2 %%%%%%%%%%%%%%%%%%%%%%%%%%%%%%%%%%%%%
 	
 	\section{  Metric and the Field Equations}
 	We consider the LRS Bianchi type I metric of the form
 	
 	\begin{equation}\label{1}
 	ds^2= -A (dx)^2 - B^2 \left( dy^2+ dz^2 \right) + dt^2
 	\end{equation}
 	Here, ``A and B" are the time-dependent metric functions. In general relativity, Einstein's field equation with the cosmological constant is:
 	
 	\begin{equation}\label{2}
 	\Lambda g_{ij}+	R_{ij}-\frac{1}{2} g_{ij} R =-T_{ij}
 \end{equation}
We consider the energy–momentum tensor in the form  $ T_{ij} = (p_{m}+\rho_{m})u_{i} u_{j}+p_{m} g_{ij} $, where $ p_{m} $ and $ \rho_{m} $ ``represents the matter pressure and matter energy density". 
%The field equations for LRS Bianchi type-I universe may be written as :
For the LRS Bianchi type-I universe, the field equations are as follows:
\begin{equation}\label{3}
2 \frac{\ddot{B}}{B}+ \frac{\dot{B}^2}{B^2}=-p_{m}+\Lambda
\end{equation}
\begin{equation}\label{4}
\frac{\ddot{A}}{A}+\frac{\ddot{B}}{B}+ \frac{\dot{A} \dot{B} }{A B}=-p_{m}+\Lambda
\end{equation}
\begin{equation}\label{5}
2\frac{\dot{A} \dot{B}}{AB}+\frac{\dot{B}^2}{B^2}=\rho_{m}+\Lambda
\end{equation}
The  volume for the model is given by  $ V =A B^{2} $, the  scale factor is assumed as $ a =\left(A B^{2}\right)^{1/3} $ and the Hubble's parameter can be written as $ H = \frac{1}{3} \left(\frac{\dot{A}}{A}+2\frac{\dot{B}}{B}\right)= \frac{\dot{a}}{a} $ in order to obtained the  solutions of the field equations.\\
From     Eq.(\ref{3})-Eq.(\ref{4}), we have
\begin{equation}\label{6}
%\frac{d}{dt}
\frac{d}{dt}\left(\frac{\dot{A}}{A}-\frac{\dot{B}}{B}\right)+\left(\frac{\dot{A}}{A}-\frac{\dot{B}}{B}\right) \left(\frac{\dot{A}}{A}+2\frac{\dot{B}}{B}\right)=0
\end{equation}
On integration, we get
\begin{equation}\label{7}
\frac{\dot{A}}{A}-\frac{\dot{B}}{B} =\frac{c_1}{a^3}
\end{equation}
Also, we have
\begin{equation}\label{8}
\frac{\dot{A}}{A}+2\frac{\dot{B}}{B} =3\frac{\dot{a}}{a}
\end{equation}
Solving Eq.(\ref{7}) and Eq.(\ref{8}), we get
\begin{equation}\label{9}
\frac{\dot{A}}{A}=\frac{\dot{a}}{a}+\frac{2}{3} \frac{c_1}{a^3}
\end{equation}
\begin{equation}\label{10}
\frac{\dot{B}}{B}=\frac{\dot{a}}{a}-\frac{1}{3} \frac{c_1}{a^3}
\end{equation}
Using Eq.(\ref{9}), Eq.(\ref{10}) in Eq.(\ref{5}), we get
\begin{equation}\label{11}
H^2 =\frac{1}{3}\left(\rho_{m}+\Lambda+\frac{1}{3} \frac{c_{1}^{2}}{a^6}\right)
\end{equation}
For barotropic matter, the energy conservation law holds i.e., $ \dot{\rho_{m}}+3 H (p_{m}+\rho_{m}) =0$. The present universe is ``dust filled for which $ p_{m}=0 $" and $ \rho_{m} \propto a^{-3} $. Using the relation $ \frac{a_{0}}{a} =1+z $ between scale factor $ a $ and redshift $ z $, we get $ \rho_{m} = \rho_{m0} (1+z)^3 $. The term $\frac{1}{3} \frac{c_{1}^{2}}{a^6} =\rho_{\sigma0} (1+z)^{6} =\rho_{\sigma}$ represents the anisotropy energy density. For the dust filled universe ($ p_{m}=0 $), the density parameters are defined as $ \Omega_{m} = \frac{\rho_{m}}{\rho_{c}}  $ and $ \Omega_{\sigma} = \frac{\rho_{\sigma}}{\rho_{c}} $, where $ \rho_{c} = \frac{3 H^{2}}{8 \pi G} $ \& $ 8 \pi G\approx 1 $. Thus, Eq.(\ref{11}) can also be witten as
\begin{equation}\label{12}
H^2 = H^{2}_{0}\left[ (1+z)^{3}\Omega_{m0}+\Omega_{\Lambda0}+ (1+z)^{6}\Omega_{\sigma0}\right]
\end{equation}

For $ z=0 $, the relationship between energy parameters is obtained from Eq.(\ref{12}) as: 
\begin{equation}\label{13}
1=\Omega_{m0} +\Omega_{\Lambda0}+\Omega_{\sigma0} 
\end{equation}

The deceleration parameter (DP) for the model can be expressed as:

\begin{equation}\label{14}
q= \frac{ (1+z)^{3}\Omega_{m0}-2\Omega_{\Lambda0}+4 (1+z)^{6}\Omega_{\sigma0}}{2 \left[ (1+z)^{3} \Omega_{m0}+\Omega_{\Lambda0}+ (1+z)^{6}\Omega_{\sigma0}\right]}
\end{equation}

The Hubble parameter $H$ and deceleration parameter $q$ is the important physical quantities for describing the evolution of the universe. For explaining the expansion of the Universe, $H$ plays a vital role and is also very useful in the estimation of the age of the universe. On the other hand, the deceleration parameter describes the phase transition (acceleration or deceleration) during the evolution of the universe.
%%%%%%%%%%%%%%%%%%%%%%%%%%%%%%%%%%%%% Section 3 %%%%%%%%%%%%%%%%%%%%%%%%%%%%%%%%%%%%%%%%%%
\section{Observasional Constraints} 

%%%%%%%%%%%%%%%%%%%%%%%%%%%%%%% Subsection 3.1 %%%%%%%%%%%%%%%%%%%%%%%%%%%%%%%%%%%%%%%%%%
\subsection{Observational Hubble Data of 46  Data Set of $ H(z) $} 
. 
In this segment, we present the observational data as well as the statistical methodological analysis that was used to constrain the model parameters of the derived Universe see in (Fig.1, 2). Here we applied 46 H(z) observational data points in the  ranges  $0\leq z\leq 2.36$, which were obtained by using the ``cosmic chronometric approach (CCA)" \cite{ref18,ref21}. \\

To determine the best-fitting values and limits for a fitted model, we use the $\chi^{2}$ statistic as 
\begin{equation}\label{15}
\chi^{2}\left(p\right)=\sum_{i=1}^{46} {\frac{\left(H_{th}(i)-H_{ob}(i)\right)^2}{\sigma(i)^2}}.
\end{equation}
Our estimated values for various cosmological parameters for minimum $\chi^{2}$ have been computed as $H_{0} = 69.48$, $\Omega_{m 0} = 0.2548$, $\Omega_{\Lambda 0} = 0.7411$, $\Omega_{\sigma 0} = 0.0005$, $\chi^{2} = 29.1136$. \\

Here $p$ is the set of model parameters, where ($p=H_{0}, \Omega_{m0}$). The $\chi^{2}$ expression in Eq. (\ref{15}) holds for the H(z) measurements listed in Table 1. This table contains a data set of the observed values of the Hubble parameters H(z) versus redshift $z$ with a possible error obtained using the different age approach by various cosmologists. 

%%%%%%%%%%%%%%%%%%%%%%%%%%%%%%%%%%%%%%%%% Table 1 %%%%%%%%%%%%%%%%%%%%%%%%%%%%%%%%%%%%%%
\begin{table}[H]
\caption{\small ``The behaviour of Hubble parameter $H(z)$ with redshift}
\begin{center}
	\begin{tabular}{|c|c|c|c|c|c|c|c|c|c|}
		\hline
		\tiny	$S.No$  &	\tiny  $Z$ & \tiny $H (Obs)$ & \tiny $\sigma_{i}$ & \tiny References & \tiny $S.No$  &	\tiny  $Z$ & \tiny $H (Obs)$ & \tiny $\sigma_{i}$ & \tiny References \\
		\hline
		\tiny	1	& \tiny 0	  &\tiny 67.77 & \tiny 1.30 & \tiny\cite{ref82} & \tiny	24	& \tiny 0.4783	 &\tiny 80.9  & \tiny9 & \tiny \cite{ref22}   \\
		
		\tiny	2	& \tiny 0.07  &\tiny 69    & \tiny 19.6 & \tiny \cite{ref83} & \tiny	25	& \tiny 0.48	 &\tiny 97 & \tiny60 & \tiny \cite{ref84}  \\

		\tiny	3	& \tiny 0.09	 &\tiny 69  & \tiny 12 & \tiny \cite{ref20}   & 	\tiny	26	& \tiny 0.51	 &\tiny 90.4  & \tiny1.9 & \tiny\cite{ref86} \\  
		%			%\hline		
		\tiny	4	& \tiny 0.01	 &\tiny 69  & \tiny 12 & \tiny \cite{ref84}   & \tiny	27	& \tiny 0.57	 &\tiny 96.8  & \tiny3.4& \tiny \cite{ref73}  \\	
		\tiny	5	& \tiny 0.12	 &\tiny 68.6  & \tiny26.2 & \tiny \cite{ref83}  & \tiny	28	& \tiny 0.593	 &\tiny 104 & \tiny 13 & \tiny \cite{ref19}  \\ 
		%			%\hline	
		%			
		\tiny	6	& \tiny 0.17	 &\tiny 83  & \tiny 8 & \tiny \cite{ref84}   & \tiny	29	& \tiny 0.60	 &\tiny 87.9  & \tiny6.1 & \tiny\cite{ref88} \\ 
		%			%\hline	
		%			
		\tiny	7	& \tiny 0.179	 &\tiny 75  & \tiny 4  & \tiny \cite{ref19}  & \tiny	30	& \tiny 0.61	 &\tiny 97.3  & \tiny2.1 & \tiny\cite{ref86} \\ 		
		%		\hline	
		\tiny	8	& \tiny 0.1993	 &\tiny 75  & \tiny 5  & \tiny\cite{ref19}   & \tiny	31	& \tiny 0.68	 &\tiny 92  & \tiny 8 & \tiny\cite{ref19}   \\ 
		%		%\hline	
		\tiny	9	& \tiny 0.2	 &\tiny 72.9  & \tiny 29.6  & \tiny \cite{ref83} & 	\tiny	32	& \tiny 0.73	 &\tiny 97.3 & \tiny 7 & \tiny \cite{ref88}   \\ 
		%		%\hline	
		\tiny	10	& \tiny 0.24	 &\tiny 79.7  & \tiny 2.7 & \tiny \cite{ref85} & \tiny	33	& \tiny 0.781	 &\tiny 105 & \tiny 12 & \tiny \cite{ref19}   \\		
		%		%\hline
		\tiny	11	& \tiny 0.27	 &\tiny 77  & \tiny 14 & \tiny \cite{ref84}   & \tiny	34	& \tiny 0.875	 &\tiny 125  & \tiny 17 & \tiny \cite{ref19}  \\ 
		%		%\hline	
		\tiny	12	& \tiny 0.28	 &\tiny 88.8  & \tiny 36.6 & \tiny \cite{ref83}  & 	\tiny	35	& \tiny 0.88	 &\tiny 90  & \tiny 40 & \tiny  \cite{ref84} \\
		%		%\hline	
		%		
		\tiny	13	& \tiny 0.35	 &\tiny 82.7 & \tiny 8.4 & \tiny \cite{ref87}    & \tiny	36	& \tiny 0.9	 &\tiny 117  & \tiny 23 & \tiny  \cite{ref84} \\ 
		%		%\hline	
		\tiny	14	& \tiny 0.352	 &\tiny 83 & \tiny 14 & \tiny \cite{ref19}    & \tiny	37	& \tiny 1.037	 &\tiny 154  & \tiny 20 & \tiny \cite{ref85} \\ 
		\tiny	15	& \tiny 0.38	 &\tiny 81.5  & \tiny 1.9 & \tiny \cite{ref86}  & 	\tiny	38 	& \tiny 1.3	 &\tiny 168 & \tiny 17 & \tiny \cite{ref84} \\ 
		%		%\hline	
		%		
		\tiny	16	& \tiny 0.3802	 &\tiny 83 & \tiny 13.5 & \tiny\cite{ref87}  & 	\tiny	39	& \tiny 1.363	 &\tiny 160  & \tiny 33.6 & \tiny \cite{ref91} \\ 
		%		%\hline	
		%		
		\tiny	17	& \tiny 0.4	 &\tiny 95  & \tiny 17 & \tiny \cite{ref20}   & \tiny	40	& \tiny 1.43	 &\tiny 177  & \tiny 18 & \tiny  \cite{ref84} \\ 
		%		%\hline	
		%		
		\tiny	18	& \tiny 0.4004	 &\tiny 77 & \tiny10.2 & \tiny\cite{ref22}   & 	\tiny	41	& \tiny 1.53	 &\tiny 140  & \tiny  14 & \tiny \cite{ref84} \\ 
		%		%\hline	
		\tiny	19	& \tiny 0.4247	 &\tiny 87.1  & \tiny 11.2 & \tiny\cite{ref22}  & 	\tiny	42	& \tiny 1.75	 &\tiny 202  & \tiny 40 & \tiny \cite{ref91}\\ 
		%		%\hline	
		\tiny	20	& \tiny 0.43	 &\tiny 86.5  & \tiny 3.7 & \tiny\cite{ref85}  & \tiny	43	& \tiny 1.965	 &\tiny 186.5  & \tiny 50.4 & \tiny \cite{ref85}  \\ 
		%		%\hline	
		\tiny	21	& \tiny 0.44	 &\tiny 82.6  & \tiny 7.8 & \tiny\cite{ref88}  & 	\tiny	44	& \tiny 2.3	 &\tiny 224 & \tiny 8 & \tiny\cite{ref16}  \\ 
		%		%\hline	
		\tiny	22	& \tiny 0.44497	 &\tiny 92.8  & \tiny 12.9 & \tiny\cite{ref22}  & 	\tiny	45	& \tiny 2.34	 &\tiny 222 & \tiny 7 & \tiny\cite{ref92} \\
		%		%\hline	
		%		
		\tiny	23	& \tiny 0.47	 &\tiny 89 & \tiny49.6  & \tiny\cite{ref89}    & 	\tiny	46	& \tiny 2.36	 &\tiny 226 & \tiny 8 & \tiny\cite{ref93}'' \\
		\hline	
	\end{tabular}
\end{center}
\end{table}
%%%%%%%%%%%%%%%%%%%%%%%%%%%%%% Subsection 3.2 %%%%%%%%%%%%%%%%%%%%%%%%%%%%%%%%%%%%%%%%%%

\subsection{Pantheon Data }
In this investigation, we use ``Pantheon data (the latest compilation of SNIa with 40 binned in the redshift range $0.014 \leq z \leq 1.62)$" \cite{ref37}. In this scenario, $\chi^2$ for Pantheon data was calculated by using the following formula.

\begin{equation}\label{16}
\chi^{2} (H_{0}, \Omega_{m0})\text{=}\sum _{i=1} \frac{(\text{$\mu $th}(\text{z }(i),H_{0}, \Omega_{m0})-\text{$\mu $obs}(i))^2}{\text{$\sigma $}(i)^2},
\end{equation}

where $\mu_{th}$ denoted the theoretical distance modulus and $\mu_{obs}$ represents
the observed ``distance modulus" with the standard error $\sigma {(i)^{2}}$. In earlier work \cite{ref38,ref40} researchers presented a technique for estimating the present values of the energy parameters $\Omega_{m0}$  $\Omega_{\Lambda_{0}}$, and $\Omega_{\sigma0}$ by ``comparing the theoretical and observed results" with
the help of the $\chi^{2}$ statistic. \\

Our estimated values for various cosmological parameters for minimum $\chi^{2}$ have been computed as $H_{0} = 70.02$, $\Omega_{m 0} = 0.2728$, $\Omega_{\Lambda 0} = 0.7262$, $\Omega_{\sigma 0} = 0.001$, $\chi^{2} = 562.242$. \\

We have formed a new data set by joining OHD and Pantheom datas and estimated an other set of values for these parameters with minimum $\chi^{2}$. In this case we obtaine $H_{0} = 70.17$, $\Omega_{m 0} = 0.2542$, $\Omega_{\Lambda 0} = 0.745731$, $\Omega_{\sigma 0} = 0.000069$, $\chi^{2} = 593.356$. \\

%%%%%%%%%%%%%%%%%%%%%%%%%%%%%%%% Fig 1a, 1b, & 1c %%%%%%%%%%%%%%%%%%%%%%%%%%%%%%%%%%
\begin{figure}[H]
\centering
(a)\includegraphics[width=7cm,height=5cm,angle=0]{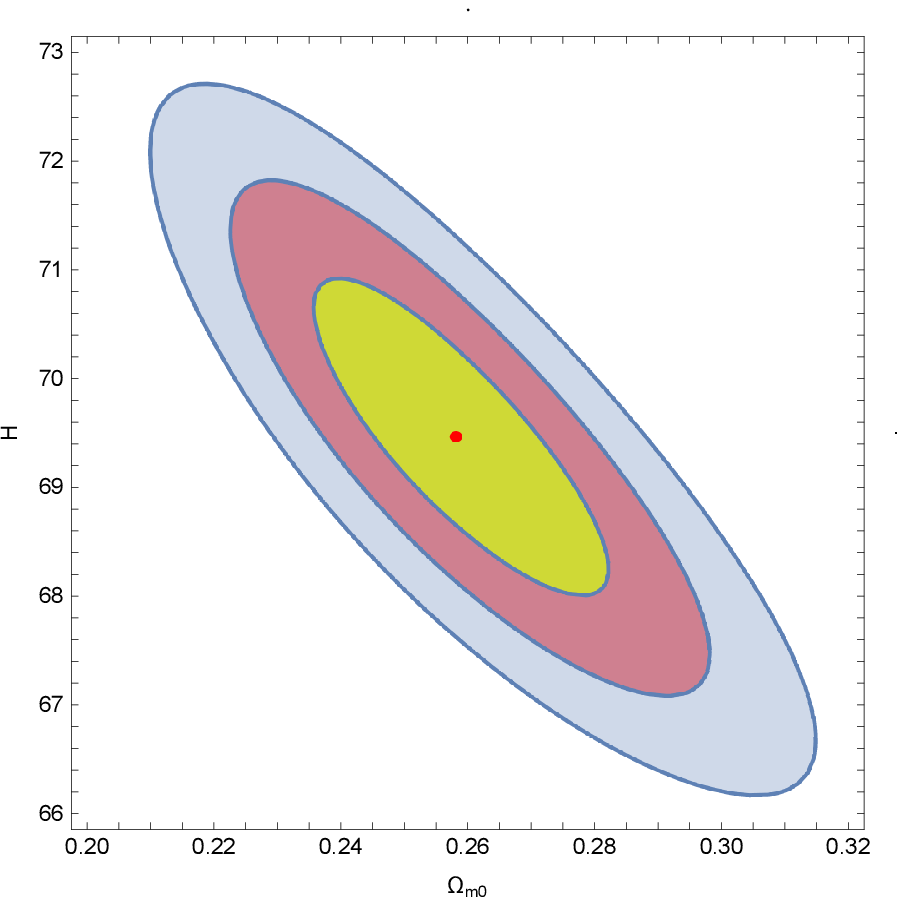}
(b)\includegraphics[width=7cm,height=5cm,angle=0]{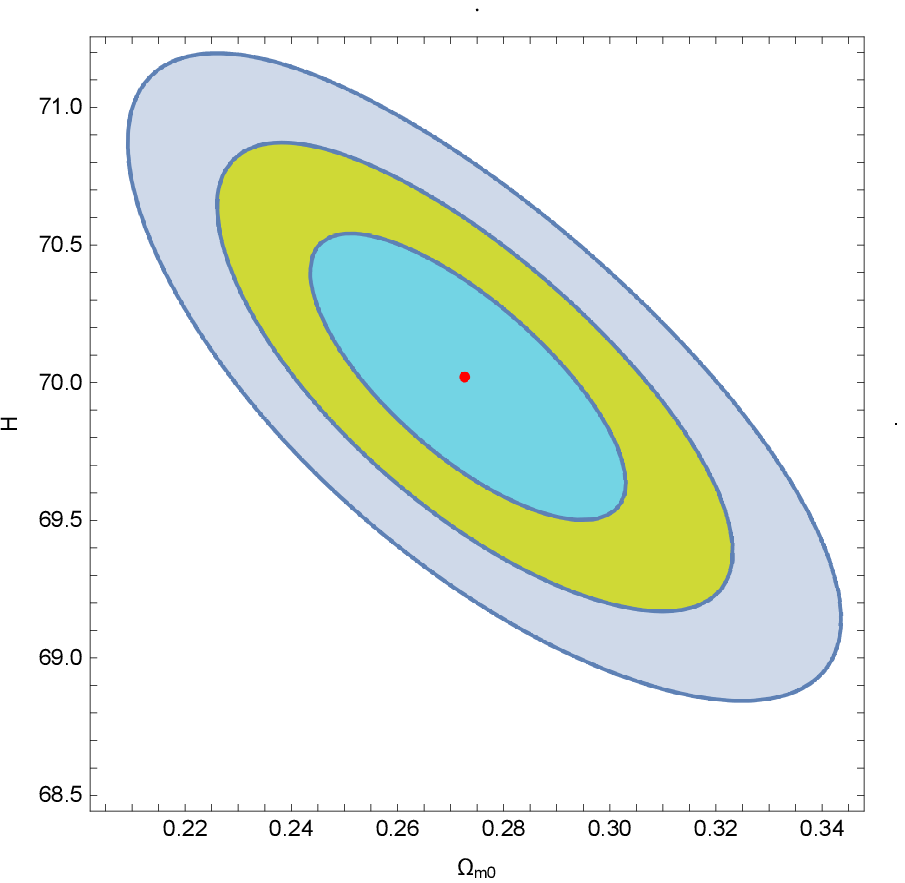}
(c)\includegraphics[width=7cm,height=5cm,angle=0]{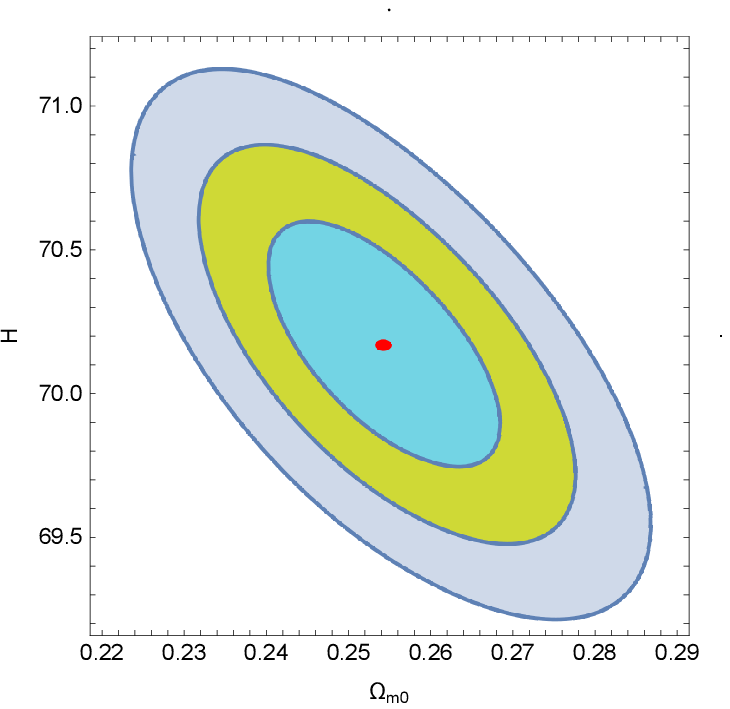}

\caption{ Two-dimensional contours for (OHD, Pantheon, OHD+Pantheon data set ) at $1\sigma$,
	$2\sigma$ and $3\sigma$ confidence regions, bounded with latest 46 OHD and Pantheon data.}\label{fig1}	
\end{figure}
%%%%%%%%%%%%%%%%%%%%%%%%%%%%%%%%%%%%%%%%%%%%%%%%%%%%%%%%%%%%%%%%%%%%%%%%%%%%%%%%%%%%%%%

%%%%%%%%%%%%%%%%%%%%%%%%% Fig 2a & 2b %%%%%%%%%%%%%%%%%%%%%%%%%%%%%%%%%%%%%%%%%%%%%%%%%%%
\begin{figure}[H]
\centering
(a)\includegraphics[width=7cm,height=5cm,angle=0]{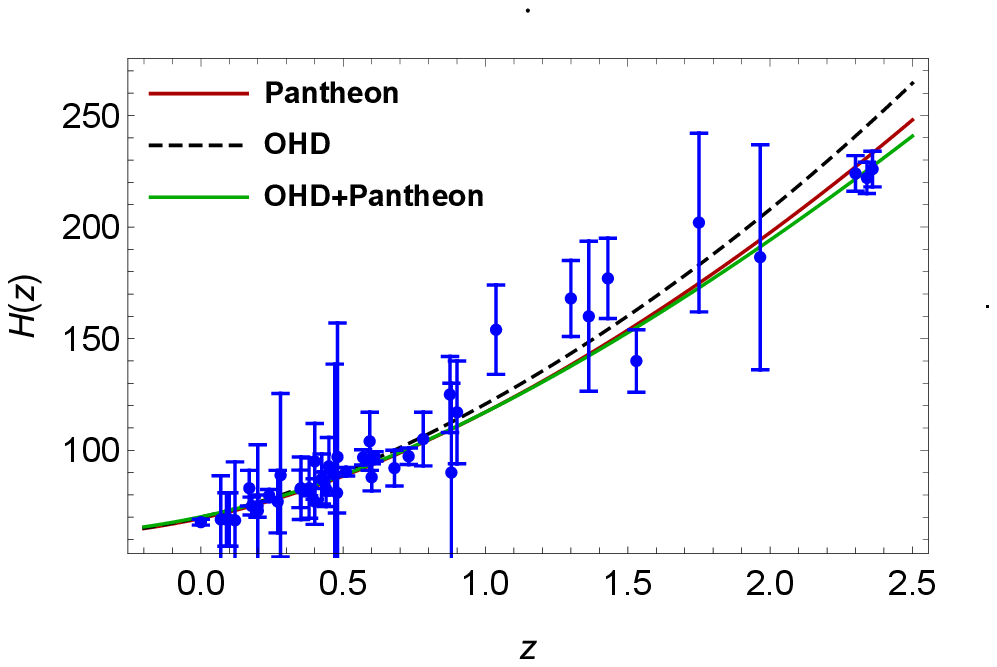}
(b)\includegraphics[width=7cm,height=5cm,angle=0]{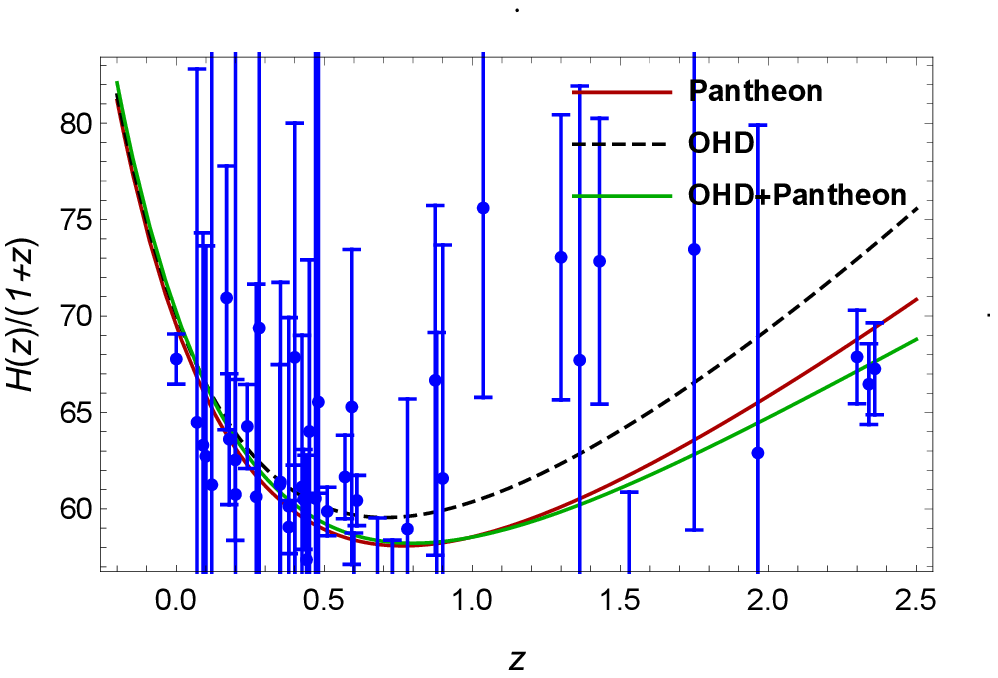}
\caption{Hubble parameter versus redshift error bar plot. The plot of Hubble rate H(z)/(1+z) versus $z$}. 
\end{figure}
%%%%%%%%%%%%%%%%%%%%%%%%%%%%%%%%%%%%%%%%%%%%%%%%%%%%%%%%%%%%%%%%%%%%%%%%%%%%%%%%%%%%%%%%%%

The Hubble parameter is clearly not a constant as demonstrated in the figures. Over the redshift, it varies slowly. Using the differential age technique and galaxy clustering method, many physicists have computed the values of the Hubble constant at various redshifts. They have described a variety of observed Hubble constant  $H_{ob}$ values as well as the corrections in the range $0 \leq z \leq 2.36$. The observed and theoretical values are found to agree quite well, indicating that our model is stable. We can see in Fig 2a that H grows as the redshift increases. The dots in the figures represents 46 observed Hubble constant $H_{ob}$ values with corrections, whereas the linear curve represents the theoretical graphs of the Hubble constant. Fig2b represents, the error bar plot of Hubble rate $H(z)/(1+z)$ versus the redshift $z$. It is clear that using a joint dataset gives raise to a better fit of the data. The dots with bars indicate the experimental observations of H(z) data, $ H_{0}$ is the present value of the Hubble constant.

%%%%%%%%%%%%%%%%%%%%%%%%  SubSubSection 3.2.1 %%%%%%%%%%%%%%%%%%%%%%%%%%%%%%%%%%%%%%%

\subsubsection{Luminosity Distance}

The luminosity distance relation  \cite{ref94,ref95} is a useful method for examining the evolution of the universe. The equation for luminosity distance $(D_{L})$ is defined in terms of redshift, as the light coming out of a distant luminous
body gets red-shifted as the cosmos expands. The luminosity distance is used to calculate a source's flux. It is  written as
\begin{equation}\label{17}
D_{L}= a_{0} (1+z) r
\end{equation}

Here  $r$ is denoted as the radial co-ordinate of the source.
The luminosity distance can be expressed as [26].
\begin{equation}\label{18}
D_{L}= (1+z) a_{0} c\int_{0}^{z}\frac{dz}{H(z)}
\end{equation}
%the  $H$ can be substituted from Eq. (43).
We noticed  that the luminosity distance is an increasing
function of redshift. 

%%%%%%%%%%%%%%%%%%%%%%%%  SubSubSection 3.2.2 %%%%%%%%%%%%%%%%%%%%%%%%%%%%%%%%%%%%%%%

\subsubsection{Apparent Magnitude and Distance Modulus in the model}
The luminosity distance is related to the distance modulus by the formula below.

\begin{equation}\label{19}
\mu = M- m_{b}= 5log_{10}\frac{D_{L}}{Mpc}+25
\end{equation}
where ``$m_{b}$ and M are the apparent and absolute magnitude of the source",
respectively. \\

We use the equation to calculate the $D_{L}$ for a supernova at very small redshift.
\begin{equation}\label{20}
D_{L} =\frac{cz}{H_{0}}
\end{equation}

There are many supernovae with low red shift whose ``apparent magnitudes" are known in the literature. The common ``absolute magnitude (M)" for Pantheon data is determined by this. Using Eqs. (\ref{19}) and (\ref{20}) in our previous study \cite{ref96,ref97}, we derived (M) as follows:
\begin{equation}\label{21}
M = 5log_{10}\frac{H_{0}}{.026c}-8.92
\end{equation}

We obtained  the expression for the apparent magnitude by using Eqs. (\ref{18}) and (\ref{19}) as:
\begin{equation}\label{22}
m_{b} = 16.08+ 5log_{10}[\frac{1+z}{.026c}\int_{0}^{z}\frac{dz}{h(z)}]
\end{equation}

We are  numerically solving Eqs.(\ref{12}) \& (\ref{21}) and by using
Pantheon data (the latest compilation of SN Ia with 40 binned in the redshift range $0.014 \leq z \leq 1.62 $. According to our model, we derive the corresponding theoretical values. Figures 3  show the closeness of observational and theoretical results, demonstrating the validity of our model.

%%%%%%%%%%%%%%%%%%%%%%%%%%%%%%%%%%% Fig 3 %%%%%%%%%%%%%%%%%%%%%%%%%%%%%%%%%%%%%%%%%%%%
\begin{figure}[H]
	\centering
	\includegraphics[scale=0.9]{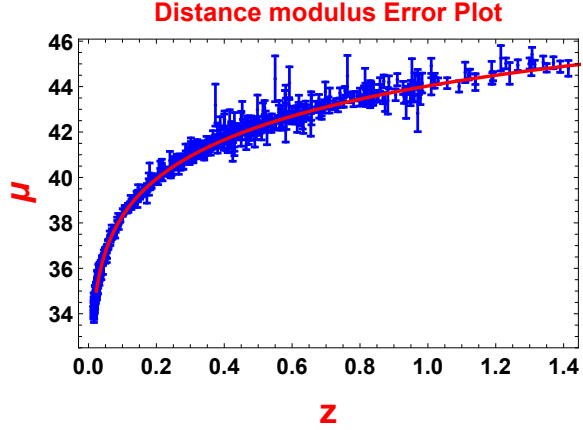}
	\caption{Plot of Distance modulus $\mu$ versus $z$}. 
\end{figure}
%%%%%%%%%%%%%%%%%%%%%%%%%%%%%%%%%%%%%%%%%%%%%%%%%%%%%%%%%%%%%%%%%%%%%%%

Figure 3 shows the progression of $\mu(z)$ for the best-fit values of model parameters Here, $\mu(z)$ represents the distance modulus, which is the difference between the ``apparent magnitude and the absolute magnitude" of the observed supernova, is given by \cite{ref89,ref90} $\mu (z) = 25 + 5log10(dL/Mpc)$, where $d_{L}$ is the luminosity distance. The blue dots in this plot correspond to the Pantheon data (the latest compilation of SNIa with 40 binned in the redshift range $0.014 \leq z \leq 1.62)$ for error bar plot  \cite{ref98}.

%%%%%%%%%%%%%%%%%%%%%%%%%%%%%%%%%%%% Section 4 %%%%%%%%%%%%%%%%%%%%%%%%%%%%%%%%%%%%%%%%%%%

\section{Age of the Universe}

The age of the cosmos is calculated as
\begin{equation}\label{23}
dt=-\frac{dz}{(1+z) H(z)}\implies \int_{t}^{t_{0}} dt=-\int_{z}^{0} \frac{1}{(1+z) H(z)} dz 
\end{equation}
Using Eq.(\ref{12}) and Eq.(\ref{23}), we get
\begin{equation}\label{24}
t_{0}-t=\int_0^z \frac{1}{H_{0} (z+1) \sqrt{\Omega_{\sigma0} (z+1)^6+\Omega_{m0} (z+1)^3+\Omega_{\Lambda0}}} \, dz
\end{equation}
The present age of universe is represents as $ t_{0} $ and it is written by
\begin{equation}\label{25}
t_{0}=\lim_{x \to \infty}\int_0^x \frac{1}{H_{0} (z+1) \sqrt{\Omega_{\sigma0} (z+1)^6+\Omega_{m0} (z+1)^3+\Omega_{\Lambda0}}} \, dz
\end{equation}
Integrating Eq.(\ref{25}), we get
\begin{equation}\label{26}
H_{0} \ t_{0} = 0.97786
\end{equation}
In this paper, we have estimated the numerical value of $ H_{0} $ as $ 0.6948 $ $ Gyr^{-1}$ $ \sim $ $69.48$ $ km s^{-1} Mpc^{-1} $. Therefore, the present age of universe for the derived model is estimated as $ t_{0}=\frac{0.97786}{H_{0}}=13.79 $ $ Gyrs $. \\

Figure 4 depicts the fluctuation of $ H_{0} (t_{0}-t) $ as a function of redshift $ z $.
According to WMAP data, the empirical value of the universe's current age is $ t_{0}=13.73^{+.13}_{-.17} $.

%%%%%%%%%%%%%%%%%%%%%%%%%%%%%%%%%%%%%%%%%% Figure 4 %%%%%%%%%%%%%%%%%%%%%%%%%%%%%%%%%
\begin{figure}[H]
\centering
\includegraphics[scale=0.9]{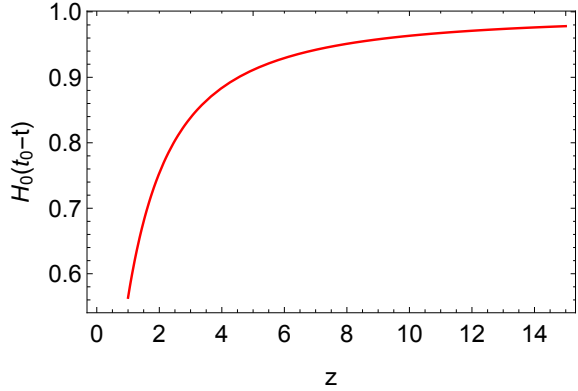}
\caption{Plot of $ H_{0} (t_{0}-t) $ versus $ z $. }\label{fig4}	
\end{figure}
%%%%%%%%%%%%%%%%%%%%%%%%%%%%%%%%%%%%%%%%%%%%%%%%%%%%%%%%%%%%%%%%%%%%%%%%%%%%%%%%%%%%%%%%

Figure 4 shows, the plot of the age of the universe versus redshift in our derived model. 
 In several cosmological research, the age of the universe is estimated as $14.46\pm0.8$ Gyrs \cite{ref99}, $14.3\pm0.6$ Gyrs \cite{ref100} and $14.5\pm1.5$ Gyrs \cite{ref101}.

%%%%%%%%%%%%%%%%%%%%%%%%%%%%%%%%%%%% Section 5 %%%%%%%%%%%%%%%%%%%%%%%%%%%%%%%%%%%%%
\section{Deceleration Parameter}

Figure 5 depicts the dynamics of the decelerating parameter $(q)$ with respect to $z$ for  ``OHD, Pantheon, and OHD + Pantheon". As we know the early universe was in decelerated era $q$, with the signature flipping at $z_{t}=0.809$ for (OHD+Pantheon), at $z_{t}=0.781$ for (OHD), at $z_{t}=0.705$ for (Pantheon), due to the dominance of dark energy in the universe. As a result, the current universe evolves with a negative sign of $q$, causing the accelerated expansion of the universe. In this way, our developed model depicts a  transition from the early deceleration phase to the current speeding phase.  Here $q_{0}= -0.42$ is the current value of the deceleration parameter. Furthermore, we find that the deceleration parameter will remain negative in the future, $z\to-1$, $q\to-1$.

%%%%%%%%%%%%%%%%%%%%%%%%%%%%%%%%%%%%%%%%%%%%% Figure 5 %%%%%%%%%%%%%%%%%%%%%%%%%%%%%%%%
\begin{figure}[H]
\centering
\includegraphics[scale=0.9]{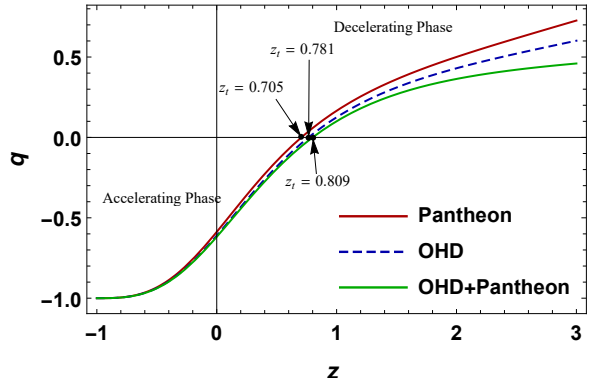}
\caption{Plot of $ q $ versus $ z $. }\label{fig5}	
\end{figure}
%%%%%%%%%%%%%%%%%%%%%%%%%%%%%%%%%%%%%%%%%%%%%%%%%%%%%%%%%%%%%%%%%%%%%%%%%%%%%%%%%%%%%%

%%%%%%%%%%%%%%%%%%%%%%%%%%%%%%%%%%%% Section 6 %%%%%%%%%%%%%%%%%%%%%%%%%%%%%%%%%%%%%%%%%%%
\section{Statefinders}

In this section, we focus on the diagnosis of the statefinder. The Hubble parameter H, which represents the universe's expansion rate, and the deceleration parameter $q$, which represents the rate of acceleration/deceleration of the expanding cosmos, are two well-known geometrical variables that characterize the universe's expansion history. They only depend on the scaling factor $a$. However, with the enhancing amount of cosmological models and the remarkable increase in the accuracies of cosmological observational data, these two parameters are no longer sensitive enough to discriminate between different models. As a consequence, the statefinder diagnosis was developed to distinguish between an increasing number of cosmological models containing dark energy. Since different cosmological models have different evolutionary paths in the $(r - ~s)$ plane, the statefinder diagnostic is likely a good way distinguish cosmological models.The remarkable property is that ``$(r-~~ s) = (1~ ~0)$" corresponds to the $\Lambda CDM$ model see in figure 6a. \\

There are so many dark energy models,  for example, the quintessence, the phantom, the Chaplygin gas, the holographic dark energy models, and the interacting DE models, which have been studied in the literature \cite{ref102,ref103,ref104,ref105,ref106}, one can clearly identify the distance from a given cosmological model to a $\Lambda CDM$ model in the $(r - ~s)$ plane.
\begin{equation}\label{27}
r= \frac{\ddot{H}}{H^{3}}+3\frac{\dot{H}}{H^2}+1=\frac{\Omega_{m0} (1+z)^{3}+\Omega_{\Lambda0}+10 \Omega_{\sigma0} (1+z)^{6}}{\Omega_{m0} (1+z)^{3}+\Omega_{\Lambda0}+\Omega_{\sigma0} (1+z)^{6}}
\end{equation}
\begin{equation}\label{28}
s= \frac{r-1}{3 (q-\frac{1}{2})}=\frac{2 \Omega_{\sigma0} (1+z)^{6}}{-\Omega_{\Lambda0}+\Omega_{\sigma0} (1+z)^{6}}
\end{equation}
%It deserves to mention here that different combinations of $r$ and $s$ represent different
%DE models\cite{ref66,ref70}.
It's worth noting that different combinations of $r$ and $s$ indicate different DE models, as referenced in \cite{ref102,ref107,ref108}. Figure 6(a) dipicts that our model lies in  quintessence region $(r < 1, s > 0)$ and  CG region $(r > 1, s < 0)$ and also meets SCDM  $(r = 1, s = 1)$ point and  LCDM  $(r = 1, s = 0)$ point.

%%%%%%%%%%%%%%%%%%%%%%%%%%%%%%%%%% Fig 6a & Fig 6b %%%%%%%%%%%%%%%%%%%%%%%%%%%%%%%%%%%%%%%%%%%
\begin{figure}[H]
\centering
(a)	\includegraphics[scale=0.90]{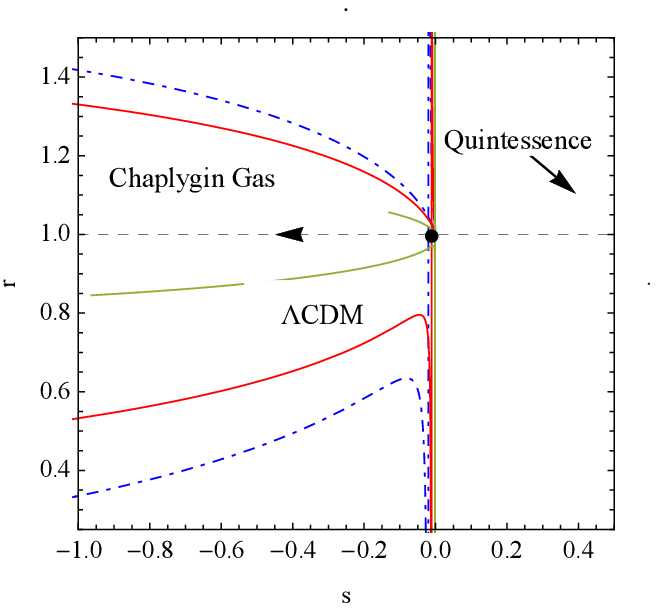}	
(b) \includegraphics[scale=0.90]{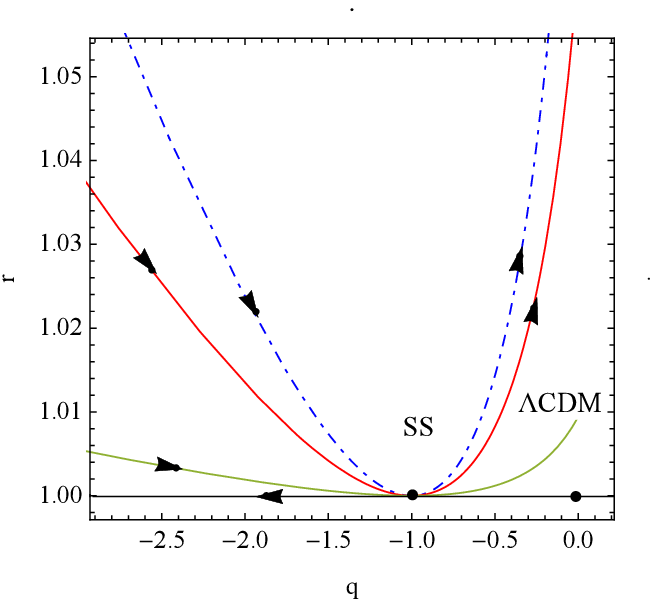}
	 \caption{(a) Plot of $ r $ vs $ s $. (b) Plot of $ r $ vs $ q $. }\label{fig6}	
\end{figure}
%%%%%%%%%%%%%%%%%%%%%%%%%%%%%%%%%%%%%%%%%%%%%%%%%%%%%%%%%%%%%%%%%%%%%%%%%%%%%%%%%%%%%%

Figure 6(b) shows the evolutionary trajectories in the  $(r-q)$ plane. In the figure the fixed point $(q =-1, r=1)$ shows the SS model (de sitter expansion) and the horizontal line represents the evolution of the trajectory corresponding to $\Lambda CDM$ $(r = 1, q = 0)$. The trajectories corresponding to our model start from the fixed point $\Lambda CDM$ and passes through the SS point.

%%%%%%%%%%%%%%%%%%%%%%%%%%%%%%%%% Section 7 %%%%%%%%%%%%%%%%%%%%%%%%%%%%%%%%%%%%%%%%%%%%
\section{Conclusion}

With the currently available data points, we have analyzed the Pantheon and OHD data and constrain several cosmological parameters. We would like to point out that our research does not take the effects of correlation. The goal of this study was to concentrate on a few key theoretical issues that have received insufficient attention in the literature. The  46 OHD data points and Pantheon data (the latest compilation of SNIa with 40 binned in the redshift range $0.014 \leq z \leq 1.62)$ were used in this analysis. In particular, we have examined the estimated values of cosmological parameters from the OHD+Pantheon data sets. While the parameter estimation errors decreased dramatically for all the data sets.\\

We have summarise the main findings once more:
\begin{itemize}
\item 
Fig.1 shows the 2-dimensional joint contours (OHD+Pantheon), at 68$\% $and 95$\% $ confidence regions,  bounded with latest . 

\item 
Fig.2 represents the error bar plots for  OHD and  Pantheon data sets. In both graphs, the dots represent observed values with corrections, whereas red lines are the present model compared with the $\Lambda CDM$ model shown in black dashed lines.

\item 
Figure 3 shows the progression of $\mu(z)$ for the best-fit values of model parameters
\item 
Our model is based on latest observational findings of 46 OHD data and Pantheon data (the latest compilation of SNIa with 40 binned in the redshift range $0.014 \leq z \leq 1.62)$. Figure 4 depicts the age- redshift plot of the universe in our derived model. In this paper, we have estimated the numerical value of $ H_{0} $ as $ 0.6948 $ $ Gyr^{-1}$ $ \sim $ $69.48$ $ km s^{-1} Mpc^{-1} $. Therefore, the present age of universe for the derived model is estimated as $ t_{0}=\frac{0.97786}{H_{0}}=13.79 $ $ Gyrs $. 

\item 
The deceleration parameter $q$ fluctuates with $z$ from positive to negative, as seen in the graph. This shows a transition from early slowdown to the current acceleration of the universe (see fig 5).

\item 
We also use the $(r,s)$ and $r,q$ planes parameters to diagnose the DE model geometrically, as shown in fig.6.Our derived model initially shows a Chaplygin gas (CG) type DE model, which later on evolves into a quintessence DE model at a few points. The model, later on, reverts back again in CG. Interestingly, the model deviates significantly from the point $(r,s)= (1,0)$ and it does not coincide with $\Lambda CDM$ (see Fig. 6).
\item
The estimated results on the basis of $\chi^{2}$ statistic using OHD, Pantheom, and OHD+ Pantheom datas are shown in the following Table-2:

%%%%%%%%%%%%%%%%%%%%%%%%%%%%%%%%% Table 2 %%%%%%%%%%%%%%%%%%%%%%%%%%%%%%%%%%%%%%%%%

\begin{table}[H]
	\caption{\small Numerical estimations for the derived models}
	\begin{center}
		\begin{tabular}{|c|c|c|c|c|c|c|}
			\hline
			\tiny	Datasets &	\tiny $H_{0}$ & \tiny $\Omega_{m0}$ & \tiny $\Omega_{\Lambda0}$ & \tiny $\Omega_{\sigma0}$ & \tiny $\chi^2$ & \tiny $z_{t}$\\
			\hline
			\tiny	$OHD$	& \tiny $69.48$	  &\tiny $0.2584$ & \tiny $0.7411$ & \tiny $0.0005$ & \tiny $29.1136 $ & \tiny $0.781 $\\
			\hline
			\tiny	$Pantheon$	& \tiny $70.02$	  &\tiny $0.2728$ & \tiny $0.7262$ & \tiny $0.001$ & \tiny $562.242 $ & \tiny $0.705 $\\
			
			\hline
			\tiny	$OHD+Pantheon$	& \tiny $70.17$	  &\tiny $0.2542$ & \tiny $0.745731$ & \tiny $0.000069$ & \tiny $593.356 $ & \tiny $0.809 $\\
			\hline	
		\end{tabular}
	\end{center}
\end{table}
%%%%%%%%%%%%%%%%%%%%%%%%%%%%%%%%%%%%%%%%%%%%%%%%%%%%%%%%%%%%%%%%%%%%%%%%%%%%%%%%%%%%%%%%

\end{itemize}
Finally, we can state that all of the preceding conclusions in LRS Bianchi-I are good and consistent with recent cosmological observations.

\section*{CRediT authorship contribution statement}
{\bf Vinod Kumar Bhardwaj} : Conceptualization, Ideas, Formulation, Writing – original draft.
{\bf Archana Dixit}: Writing-review \& editing, Conceptualization. {\bf Rita Rani}: Analysis of figures and review of literatures.
 {\bf G. K. Goswami}: Formal analysis, Writing – review \& editing.
 {\bf Anirudh Pradhan}: Methodology, Supervision, Final writing \& editing.

\section*{Declaration of competing interest}
The authors declare that they have no known competing financial interests or personal relationships that could have appeared to influence the work reported in this paper.
%%%%%%%%%%%%%%%%%%%%%%%%%%%%%%%%%%%%%%%%%%%%%%%%%%%%%%%%%%%%%%%%%%%%%%
\section*{Acknowledgement}
A. Pradhan thanks to the IUCAA, Pune, India for providing support and facility under associateship program.
%%%%%%%%%%%%%%%%%%%%%%%%%%%%%%%%%%%%%%%%%%%%%%%%%%%%%%%%%%%%%%%%%%%%%%%%%%%%%%%%%%%%%%%%%%%%%%


\begin{thebibliography}{99}
	\bibitem {ref1}
	
	S. Perlmutter, G. Aldering $\&$ M. Valle {\it et al.}, Discovery of a supernova explosion at half the age of the Universe and its cosmological implications, {\it Nature}  {\bf 391} (1998) 51-54.
	
	\bibitem {ref2}
	S. Perlmutter {\it et al.}, Measurements of $\Omega$ and $\Lambda$ from 42 high-redshift supernovae, {\it  Astrophysical J } {\bf 517} (1999) 565.
	\bibitem {ref3}
	A. G. Riess {\it et al.}, Type Ia supernova discoveries at $z> 1$ from the Hubble Space Telescope: Evidence for past deceleration and constraints on dark energy evolution,{\it APJ} {\bf 607}  (2004) 665.
	
	
	\bibitem {ref4}
	W. Baade  $\&$ F.Zwicky, Photographic Light-Curves of the Two Supernovae in IC 4182 and NGC 1003, {\it The Astrophys J.} {\bf 88} (1938) 411.
	

	
	\bibitem {ref5}
	S. A. Colgate, Supernovae as a standard candle for cosmology, {\it Astrophys. J.} {\bf 232} (1979) 404.
	
	\bibitem {ref6}
	A. Goobar $\&$  S. Perlmutter, Feasibility of measuring the cosmological constant Lambda and mass density Omega using type Ia supernovae, {\it Astrophys. J.} {\bf 450} (1995) 14. arXiv preprint astro-ph/9505022.
	
	\bibitem {ref7}
	S. Perlmutter, G. Aldering $\&$ M. Valle {\it et al.}, Discovery of a supernova explosion at half the age of the Universe, {\it Nature}  {\bf 391} (1998) 51-54.
	
	\bibitem {ref8}
	A. G. Riess {\it et. al.}, Observational evidence from supernovae for an accelerating universe and a cosmological constant, {\it Astrophys. J.} {\bf 116} (1998) 1009.
	
	\bibitem {ref9}
	P. A. R. Ade  {\it et al.}, Planck 2013 results XVI Cosmological parameters,  {\it Astron. Astrophys.} {\bf 571(A16)} (2014), arXiv:1303.5076 [astro-ph.CO]
	
	\bibitem {ref10}
	H. Campbell {\it  et al.}, Cosmology with photometrically classified type ia supernovae from the sdss-II supernova survey, {\it Astrophys. J.} {\bf 763(2)} (2013) 88.
	
	\bibitem {ref11}
	V. Salzano {\it et al.}, Linear dark energy equation of state revealed by supernovae, {\it Mod. Phys. Lett. A} {\bf 29(2)} (2014) 1450008.
	
	
	\bibitem {ref12}
	F. Beutler {\it et al.}, The 6dF Galaxy Survey: baryon acoustic oscillations and the local Hubble constant,  {\it Mon. Not. R. Astron. Soc.} {\bf 416(4)} (2011) 3017.
	
	\bibitem {ref13}
	C. Blake {\it et al.}, The Wigglez Dark Energy Survey: mapping the distance–redshift relation with baryon acoustic oscillations,  {\it Mon. Not. R. Astron. Soc.} {\bf 418} (2011) 1707.
	
	\bibitem {ref14}
	K. T. Mehta  {\it et al.}, A 2 percent distance to $z= 0.35$ by reconstructing baryon acoustic oscillations-III. Cosmological measurements and interpretation, {\it Mon. Not. R. Astron. Soc.} {\bf 427} (2012) 2168.
	
	\bibitem {ref15}
	W. J. Percival {\it et al.}, Baryon acoustic oscillations in the Sloan Digital Sky Survey data release 7 galaxy sample, {\it Mon. Not. R. Astron. Soc.} {\bf 401} (2010) 2148. 
	
	\bibitem {ref16}
	N. G. Busa {\it et al.}, Baryon acoustic oscillations in the Ly$\alpha$ forest of BOSS quasars, {\it Astron $\&$ Astrophys} {\bf 552} (2013) A96.
	%N. G. Busca et al., Baryon acoustic oscillations in the Ly forest of BOSS quasars, {\it Astron. Astrophys.} {\bf 552(A96)} (2013). 
	
	\bibitem {ref17}
	Y. Chen and B. Ratra, Hubble parameter data constraints on dark energy, {\it Phys. Lett. B} {\bf 703(4)} (2011) 406.
	
	\bibitem {ref18}
	O. Farooq, B. Ratra, Hubble parameter measurement constraints on the cosmological deceleration-acceleration transition redshift, {\it Astrophys. J. Lett.} {\bf 766} (2013) L7.
	
	
	\bibitem {ref19}
	M. Moresco {\it et al.}, Improved constraints on the expansion rate of the Universe up to $z\sim 1.1$ from the spectroscopic evolution of cosmic chronometers, {\it  J. Cosmol. Astropart. Phys.} {\bf 2012} (2012) 006.
	
	
	\bibitem {ref20}
	J. Simon, L. Verde, $\&$  R. Jimenez, Constraints on the redshift dependence of the dark energy potential, {\it Phys. Rev. D} {\bf 71} (2005) 123001.  
	
	
	\bibitem {ref21}
	O. Farooq, Observational constraint on dark energy cosmological model parameters,{\it Kansas State University PhD Thesis} [arXiv:1309.3710] 2013.
	
	\bibitem {ref22}
	M. Moresco et al., A 6 $\%$ measurement of the Hubble parameter at z $\sim 0.45$ direct evidence of the epoch of cosmic re-acceleration, {\it J. Cosmol. Astropart. Phys.}  {\bf 05} (2016) 014 . 
	
	\bibitem {ref23}  
	O. Akarsu, T. Dereli and S. Kumar,  Probing kinematics and fate of the Universe with linearly time-varying deceleration parameter, {\it Eur. Phys. J. Plus} {\bf 129} 2014, 1.
	
	\bibitem {ref24}  
	L. Samushia and B. Ratra, Cosmological constraints from Hubble parameter versus redshift data,  {\it Astrophys. J.}  {\bf 650} 2006 L5.
	
	\bibitem {ref25}
	L. P. Chimento and M. G. Richarte,  Nonbaryonic dark matter and scalar field coupled with a transversal interaction plus decoupled radiation, {\it Eur. Phys. J. Plus}, {\bf 73}  2013, 1.
	
	\bibitem {ref26}
	C. Gruber and O. Luongo, Cosmographic analysis of the equation of state of the universe through Padé approximations, {\it Phys. Rev. D} {\bf 89(10)} (2014) 103506.
	
	\bibitem {ref27}
	K. Bamba, Cosmological investigations of (extended) nonlinear massive gravity schemes with nonminimal coupling, {\it Phys. Rev. D} {\bf 89} (2014) 083518.
	
	\bibitem {ref28}
	M. Forte, On extended sign-changeable interactions in the dark sector, {\it  Gen. Rel. Grav.}, {\bf 46}  (2014) 1811. 
	
	\bibitem {ref29}
	S. Capozziello, O. Farooq, O. Luongo,  Cosmographic bounds on the cosmological deceleration-acceleration transition redshift in $ f (\mathcal {R}) $ gravity, {\it Phys.Rev. D} {\bf 90} (2014) 044016.
	
	\bibitem {ref30}
	R. Y. Guo  and X. Zhang, Constraining dark energy with Hubble parameter measurements: an analysis including future redshift-drift observations, { \it Eur. Phys. J. C}  {\bf76}  (2016) 163.
	
	\bibitem {ref31}
	L. Verde, P. Protopapas $\&$ R. Jimenez, The expansion rate of the intermediate universe in light of Planck, { \it Phys.Dark Univ.} { \bf 5} (2014) 307.
	
	\bibitem {ref32}
	Y. Chen, S. Kumar $ \&$ B. Ratra,  Determining the Hubble constant from Hubble parameter measurements, {\it Astrophys J.} {\bf 835} (2017) 86.
	\bibitem {ref33}
	W. Wilson and M. O Farooq, Using Hubble Parameter Measurements to Find Constraints on Dark Energy Based on Different Cosmological Models, {\it J. Appl. Math. Comput.} {\bf 1} (2017) 1.
		\bibitem {ref34}
	O. Farooq {\it et al.},  Hubble parameter measurement constraints on the redshift of the deceleration–acceleration transition, dynamical dark energy, and space curvature,  {\it Astrophys J.} {\bf 835} (2017) 26.
	
	\bibitem {ref35}
	G. K. Goswami {\it et al.}, Modeling of Accelerating Universe with Bulk Viscous Fluid in Bianchi V Space‐Time, {\it Fortschritte der Physik} {\bf 69(6)} (2021) 2100007.
	
	\bibitem {ref36}
	P. Shrivastava {\it et al.}, The simplest parametrization of equation of state parameter in the scalar field Universe, [ arXiv:2107.05044].
	
	\bibitem {ref37}
	D. M. Scolnic {\it et al.}, The complete light-curve sample of spectroscopically confirmed SNe Ia from Pan-STARRS1 and cosmological constraints from the combined pantheon sample, {\it Astrophys J.} {\bf 859} (2018) 101.
		\bibitem {ref38}
	G. K. Goswami, A. K. Yadav $\&$ B. Mishra, Probing kinematics and fate of Bianchi type V Universe, {\it Mod. Phys Lett. A} {\bf 35} (2020) 2050224.
	
	\bibitem {ref39}	
	H. Amirhashchi $\&$  A.K Yadav, Interacting Dark Sectors in Anisotropic Universe: Observational Constraints and $ H_ {0} $ Tension, arXiv:2001.03775 (2020) [astroph.CO]
	
	\bibitem {ref40}	
	G. K. Goswami, M. Mishra, A. K. Yadav $\&$  A. Pradhan,  Two-fluid scenario in Bianchi type-I universe, {\it Mod. Phys Lett. A} {\bf  35} 2050086 (2020).
	
	\bibitem {ref41}	
	S. Kumar $\&$  C. P. Singh, Anisotropic dark energy models with constant deceleration parameter, {\it Gen. Relativ. Grav.} {\bf  43} (2011) 1427.

	\bibitem {ref42}		
	S. Capozziello {\it et al.}, Cosmological viability of f (R)-gravity as an ideal fluid and its compatibility with a matter dominated phase, {\it Phys. Lett. B} {\bf 639} (2006) 135..
	
%	\bibitem {ref43}
%	S. M. Carroll {\it et al.}, Is cosmic speed-up due to new gravitational physics?, {\it Phys. Rev. D} {\bf 70} (2004) 043528.	
	
	\bibitem {ref43}
	C. Aktas {\it et al.}, Behaviors of dark energy and mesonic scalar field for anisotropic universe in $f(R)$ gravity, {\it Phys Lett. B} {\bf 707} (2012) 237.
		\bibitem {ref44}
	S. Nojiri, S. D. Odintsov, Unified cosmic history in modified gravity: from F(R) theory to Lorentz non-invariant models, {\it Phys. Rep.} {\bf 505} (2011) 59.
	
	\bibitem {ref45}	
	K. Bamba {\it et al.}, Dark energy cosmology: the equivalent description via different theoretical models and cosmography tests, {\it Astrophys. Space Sci.} {\bf 342} (2012) 155.
	
	\bibitem {ref46}
	M. F. Shamir, Dark-energy cosmological models in f(G) gravity, {\it Exp. Theor. Phys.} {\bf 123(4)} (2016) 607.
	
	\bibitem {ref47}
	R. Zia, D. C.  Maurya $\&$  A. K Shukla,  Transit cosmological models in modified f (Q, T) gravity, {\it  Int. J. of Geom. Method Mod. Phys.} {\bf 18} (2021) 2150051.
	
	\bibitem {ref48}	
	A. Pradhan  $\&$ A.  Dixit, Transit cosmological models with observational constraints in f (Q, T) gravity, {\it  Int. J. of Geom. Method Mod. Phys.} {\bf 18}  (2021) 2150159.
	
	
	\bibitem {ref49}
	A. Pradhan, D. C. Maurya  $\&$ A.  Dixit, Dark energy nature of viscus universe in f(Q)-gravity with observational constraints, {\it  Int. J. of Geom. Method Mod. Phys.} {\bf 18} (2021) 2150124.
	
	\bibitem {ref50}
	N. Frusciante, Signatures of f (Q) gravity in cosmology, {\it Phys. Revi D} {\bf 103} (2021) 044021.
		\bibitem {ref51}
	Banerjee et al, Wormhole geometry in f (Q) gravity and the energy conditions, {\it Europ. Phys. J. C} {\bf 81} (2021) 10131, arXiv:2109.15105[gr-qc].
	
	\bibitem {ref52}
	S. V. Lohakare, S. K. Tripathy  $\&$  B. Mishra, Cosmological model with time varying deceleration parameter in F (R, G) gravity, {\it  Phys. Scr.} {\bf  96} (2021) 125039.
	
	\bibitem {ref53}
	M. Caruana,  G.  Farrugia  $\&$ J. Levi Said,  Cosmological bouncing solutions in f (T, B) gravity, {\it  Eur. Phys. J. C} {\bf 80} (2020) 1.
	
		\bibitem {ref54}
	T. Tangpati et al.,  Quark star in the Einstein-Gauss-Bonnet theory: a new branch of stellar configurations, {\it Annals Phys.} {\bf 430} (2021) 168498.
	
		\bibitem {ref55}
	T. Tangpati et al.,  Anisotropic quark stars in the Einstein-Gauss-Bonnet theory, {\it Phys. Lett. B} {\bf 819} (2021) 136423.
	
		\bibitem {ref56}
	T. Tangpati et al.,  Anisotropic quark stars in 4D Einstein-Gauss-Bonnet theory, {\it Phys. Dark Univ.} {\bf 33} (2021) 100877, arXiv:2109.00195[gr-qc].
	
		\bibitem {ref57}
	S. K. Maurya et al., Minimally deformed anisotropic stars by gravitaional decpoupling in Einstein-Gauss-Bonnet gravity, {\it  Eur. Phys. J. C} {\bf 81} (2021) 848.
	
		\bibitem {ref58}
	Jaun M, Z, Pretel et al., Electrically charged quark stars in 4D Einstein-Gauss-Bonnet gravity, {\it  Eur. Phys. J. C} {\bf 82} (2022) 180, arXiv:2108.07454[gr-qc].
	
	\bibitem {ref59}
	G. Panotopoulos et al., Charged polytropic compact stars in 4D Einstein-Gauss-Bonnet gravity, {\it Chin. J. Phys.} Online published on 4 Feb. 2022. https://doi.org/10.1016/j,cjph.2022.01.008.
	
	\bibitem {ref60}
	T. Harko, F. S. N. Lobo, S. Nojiri, S. D. Odintsov,  $f(R, T)$ gravity, {\it 
		Phys.Rev. D} {\bf 84} (2011)  024020.
	
	\bibitem {ref61}
	 G.P. Singh, B.K. Bishi, P.K. Sahoo, Cosmological constant $\Lambda$ in $f(R,T)$ modified gravity, Int. J. Geom. Methods Mod. Phys. 13
	(2016) 1650058.

	 \bibitem{ref62}
	 G.K. Goswami, R.N. Dewangan, A.K. Yadav, Grav. Cosmol. 22 (2016) 388.
		\bibitem {ref63}
	 G.K. Goswami, et al., Modern Phys. Lett. A 33 (2020) 2050086.
		\bibitem {ref64}
	 H. Amirhashchi, S. Amirhashchi, Phys. Dark Univ. 29 (2020) 100557.
		\bibitem {ref65}
	 U.K. Sharma, G.K. Goswami, A. Pradhan, Gravit. Cosmol. 24 (2018) (2018)
	191.
	\bibitem {ref66}
		Anil Kumar Yadav, et al., Phys. Dark Univ. 31 (2021) 100738
		\bibitem {ref67}	
	 A.G. Doroshkevich, Ya.B. Zeldovich, JETP Lett. 5 (1967) 3.
		\bibitem {ref68}
	 C.W. Misner, Phys. Rev. Lett. 19 (1967) 533.
		\bibitem {ref69}
     C.L. Bennett, et al., Astrophys. J. Suppl. Ser. 148 (2003) 1043
		\bibitem {ref70}		
	 O. Heckmann, E. Schucking, in: L. Witten (Ed.), Relativistic Cosmology in
	Gravitation: An Introduction to Current Research, Willey, New York, 1962,
	pp. 438–469, Chap XI.
		\bibitem {ref71}
	 G.F.R. Ellis, M.A.H. MacCallum, Commun. Math. Phys. 12 (1969) 108.
	
		\bibitem {ref72}		
	J. K Singh  $\&$ S. Rani, Spatially Homogeneous Bianchi Type-I Universes with Variable $G$ and $\Lambda$, {\it Int. J. Theor. Phys.} {\bf 52} (2013) 3737.

	\bibitem {ref73}
	P. Sarmah $\&$  U. D. Goswami,   Bianchi Type I model of universe with customized scale factors (2022). [arXiv:2203.00385].
	
	\bibitem {ref74}
	A. K. Yadav, P. K. Sahoo $\&$ V. Bhardwaj,  Bulk viscous Bianchi-I embedded cosmological model in $f (R, T)= f 1 (R)+ f 2 (R) f 3 (T)$ gravity, {\it Mod. Phys. Lett.A } {\bf 34} (2019)1950145.
		\bibitem {ref75} 
	A. K. Yadav, A. Pradhan,  $\&$  A. K.Singh, Bulk viscous LRS Bianchi-I Universe with variable G and decaying $\Lambda$, {\it Astrophys. Space Sci.} {\bf 337} (2012) 379.
	
	\bibitem {ref76}
	A. Pradhan, R. K. Tiwari, A. Beesham $\&$ R. Zia, LRS Bianchi type-I cosmological models with accelerated expansion in f (R, T) gravity in the presence of $\Lambda $ $\Lambda$(T), {\it Eur. Phys. J. Plus} {\bf 134} (2019) 1.
	
		\bibitem {ref77}
	A. Pradhan, V. K. Bhardwaj, A.  Dixit  $\&$ S. Krishnannair, A new class of holographic dark energy models in LRS Bianchi Type-I, {\it  Int. J Mod Phys A } {\bf 36} (2021) 2150256.
	\bibitem {ref78}
	A. Pradhan $\&$ A. Kumar, LRS Bianchi I cosmological universe models with varying cosmological term $\Lambda$,  {\it  Int. J. Mod. Phys. D }{\bf 10} (2001) 291.

	\bibitem {ref79}
	A. Pradhan $\&$ S. K.  Singh, Bianchi type I magnetofluid cosmological models with variable cosmological constant revisited,{\it  Int. J. Mod. Phys. D }{\bf 13} (2004) 503.

	
	\bibitem {ref80}
	S. Agarwal, R. K. Pandey $\& $ A.Pradhan, LRS Bianchi type II perfect fluid cosmological models in normal gauge for Lyra’s manifold,{\it Int. J. Theor. Phys.} {\bf 50}  (2011) 296.
	\bibitem {ref81}
	A. Pradhan S. Agarwal,  $\&$ G. P.  Singh,  LRS Bianchi Type-I Universe in Barber’s Second Self Creation Theory, {\it Int. J. Theor. Phys.} {\bf 48}  (2009) 158.
	

	
	 \bibitem {ref82}
	 E. Macaulay {\it et al.}, First cosmological results using Type Ia supernovae from the Dark Energy Survey: measurement of the Hubble constant, {\it Mon. Not. R. Astro. Soc.} {\bf 486} (2019) 2184.
	 \bibitem {ref83}
	 C. Zhang {\it et al.}, Four new observational H(z) data from luminous red galaxies in the sloan digital sky survey data release seven, {\it Res. Astron. Astrophys} {\bf 14} (2014) 1221.
	 
	 \bibitem {ref84}
	 D. Stern {\it et al.}, Cosmic chronometers: constraining the equation of state of dark energy.I: H(z) measurements, {\it J. Cosmol. Astropart. Phys.} {\bf 1002} (2010) 008 .
	 
	 
	 \bibitem {ref85}
	 E. Gazta Naga {\it et al.}, Clustering of luminous red galaxiesIV. Baryon acoustic peak in the line-of-sight direction and a direct measurement of H(z),{\it Mon. Not. R. Astro. Soc.} {\bf 399} (2009) 1663.
	 
	 \bibitem {ref86}
	 D. H Chauang and Y. Wang , Modelling the anisotropic two-point galaxy correlation function on small scales and single-probe measurements of H(z), DA(z) and f(z) $\sigma$8(z) from the sloan digital sky survey DR7 luminous red galaxies, {\it Mon. Not. R. Astro. Soc.} {\bf 435} (2013)255.
	 
	 \bibitem {ref87}
	 S. Alam {\it et al.}, The clustering of galaxies in the completed SDSS -III Baryon Oscillation Spectroscopic Survey: cosmological analysis of the DR12 galaxy sample, {\it Mon. Not.R.Astron. Soc.} {\bf 470} (2017) 2617.
	 
	 
	 \bibitem {ref88}
	 C. Blake {\it et al.} The Wiggle Z Dark Energy Survey: joint measurements of the expansion and growth history, {\it Mon. Not. R. Astron. Soc.} {\bf 425} (2012) 405.
	 
	 \bibitem {ref89}
	 A. L. Ratsimbazafy {\it et al.}, Age-dating luminous red galaxies observed with the Southern African Large Telescope,{\it Mon. Not. R. Astron. Soc.} {\bf 467} (2017) 3239.
	 \bibitem {ref90}
	 L. Anderson{\it  et al.}, The clustering of galaxies in the SDSS -III Baryon Oscillation Spectro-scopic Survey: baryon acoustic oscillations in the Data Releases 10 and 11 Galaxy samples, {\it Mon. Not. R. Astron. Soc.} {\bf 441} (2014) 24.
	 \bibitem {ref91}
	 M. Moresco, Raising the bar: new constraints on the Hubble parameter with cosmic chronometers at z $\equiv$ 2, {\it Mon. Not. R. Astron. Soc.} {\bf 450} (2015) L16.
	 
	 
	 \bibitem {ref92}
	 T. Delubac {\it et al.} , Baryon acoustic oscillations in the $Ly\alpha$ forest of BOSSDR11 quasars, {\it Astron $\&$ Astrophys} {\bf 574} (2015) A59.
	 
	 \bibitem {ref93}
	 A. Font-Ribera, {\it et al.}, Quasar-Lyman $\alpha$ forest cross-correlation from BOSS DR11: BaryonAcoustic Oscillations, {\it J. Cosmol. Astropart. Phys.} {\bf 2014} (2014) 027.
	 
	 \bibitem {ref94} 
	 A. R. Liddle and D. H. Lyth, Cosmological inflation and
	 large-scale structure (Cambridge University Press, Cambridge, 2000)
	 
	 \bibitem {ref95}
	 S. M. Carroll $\&$ M. Hoffman, Can the dark energy equation-of-state parameter $\omega$ be less than -1, {\it Phys. Rev. D}  (2003) 023509.
	 \bibitem {ref96}
	 G. K. Goswami, Cosmological parameters for spatially flat dust filled Universe in Brans-Dicke theory, {\it Res. Astron. Astrophys.} {\bf 17(3)} (2017) 27 .
	 
	 \bibitem {ref97}
	 G. K. Goswami, FRW Cosmological Model In Present Perspective, {\it Can. J. Phys.} {\bf 97} (2019) 588.
	 
	 \bibitem {ref98} 
	 N. Suzuki {\it et al.} The Hubble Space Telescope cluster supernova survey. Improving the dark-energy constraints above $z> 1$ and building an early-type-hosted supernova sample, {\it Astrophy. J} {\bf 746}, (2012) 85.
	 
	 
	 \bibitem {ref99}
	 H. E. Bond {\it et al.}, HD 140283: A star in the solar neighborhood that formed shortly after the Big Bang, {\it Astrophys. J.} {\bf 765} (2013) L12.
	 
	 
	 \bibitem {ref100}
	 S. Masi {\it et al.} The BOOMERanG experiment and the curvature of the Universe, {\it Prog. Part. Nucl. Phys.} {\bf  48} (2002) 243.
	 
	 \bibitem {ref101}
	 A. Renzini, A. Bragaglia $\&$ F. R.  Ferraro, The white dwarf distance to the globular cluster NGC 6752 (and its age) with the Hubble Space Telescope, {\it Astrophys. J.},{\bf  465} (1996) L23.
	 
	 \bibitem {ref102}  
	 U. Alam, V. Sahni, T. Deep Saini $\&$ A. A. Starobinsky, Exploring the expanding universe and dark energy using the Statefinder diagnostic,  {\it Mon. Not.  Roy Astron Soc.} {\bf 344} (2003) 1057.
	 
	 \bibitem {ref103}
	 
	 A.  Ujjaini, T. D. Saini, $\&$ V. Sahni, Exploring the expanding universe and dark energy using the statefinder diagnostic, (2011)
	 \bibitem {ref104}
	 X.  Zhang,  Statefinder diagnostic for holographic dark energy model,{\it Int. J.  Mod. Phys D} {\bf 14},(2005) 1597.
	 
	 \bibitem {ref105}
	 M. R. Setare, J. Zhang $\&$ X. Zhang,  Statefinder diagnosis in a non-flat universe and the holographic model of dark energy, {\it JCAP},  {\bf 3} (2007) 007.
	 
	  \bibitem {ref106}
	  U. K. Sharma $\&$ A. Pradhan, Diagnosis Tsallis holographic dark energy models with statefinder and $\omega - \omega^{'}$, {\it Mod. Phys. Lett. A}, {\bf 34} (20019) 1950101.
	 \bibitem {ref107}
	 V. Sahni,  T. D Saini, A. A. Starobinsky, U. Alam, Statefinder—A new geometrical diagnostic of dark energy, {\it J. Exp. Theor. Phys.
	 	Lett.} {\bf 77} 2003, 201.
	 \bibitem {ref108} 
	 V. K. Bhardwaj, A.  Dixit,  $\&$ A. Pradhan, Statefinder hierarchy model for the Barrow holographic dark energy, {\it New Astronomy} {\bf 88} (2021) 101623.
	 
	\end{thebibliography}
\end{document}